\documentclass[10pt,letterpaper]{article}
\IfFileExists{spconf.sty}{\usepackage{spconf}}{\usepackage{isbi_preview}}
\usepackage{amsmath,amssymb,graphicx,booktabs,microtype,cite}
\usepackage[hidelinks]{hyperref}
\usepackage[font=small]{caption}
\graphicspath{{figures/}}
\hypersetup{pdftitle={Kidney-Centered CT--Histology Cross-Attention for Renal Cancer Survival Prediction},pdfsubject={ISBI 2027 manuscript draft}}
\title{
When Slides Met Scans: Cross-Attention for Renal Cancer Prognosis}
\name{Anusha Krishnamoorthy$^{1}$, Basudha Pal$^{2}$}
\address{$^{1}$Wallace H. Coulter Department of Biomedical Engineering, Georgia Institute of Technology \& Emory University,\\$^{2}$Department of Electrical \& Computer Engineering, Johns Hopkins University}
\begin{document}
\maketitle
\begin{figure*}[t]
    \centering
    \includegraphics[width=\textwidth,height=0.3\textheight,keepaspectratio]{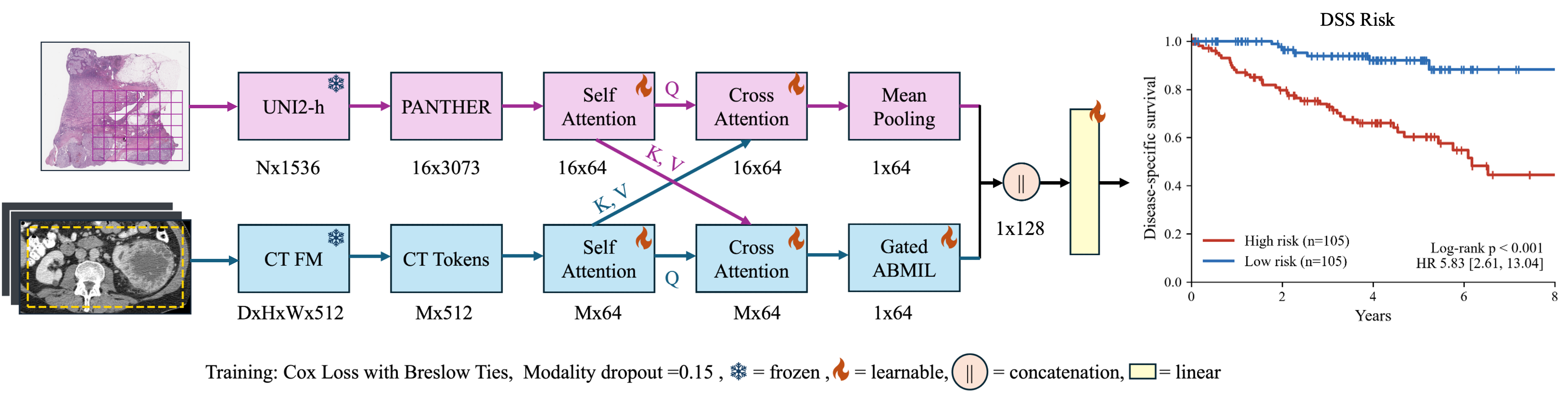}
    
    \caption{Overall architecture diagram}
    \label{fig:teaser}    
\end{figure*}
\begin{abstract}Histopathology and computed tomography (CT) provide complementary views of renal cancer. We connect these modalities through bidirectional cross-attention between 16 histology prototypes and up to 32 kidney-centered CT tokens from frozen foundation models. In 210 TCGA-KIRC patients with 46 disease-specific survival (DSS) events, 5-fold evaluation yielded a C-index of $0.760\pm0.081$ and 3-year AUROC of 0.773 (95\% CI 0.678--0.860), the highest averages among evaluated unimodal and fusion baselines. External testing in 180 CPTAC-CCRCC patients yielded DSS C-index of 0.629 with CT withheld. Attribution analysis linked predictions to image regions while exposing acquisition artifacts. These findings support further study of cross-scale fusion and independent multimodal validation.
\end{abstract}

\section{Introduction}
Renal cell carcinoma exhibits heterogeneity across microscopic tissue architecture and macroscopic tumor anatomy, motivating prognostic models that integrate information across imaging scales \cite{sparra}. Histopathology captures cellular morphology and tissue organization, while computed tomography provides a volumetric view of the tumor and surrounding anatomy. Combining these complementary perspectives offers an opportunity to characterize disease aggressiveness more comprehensively than either modality alone.

Foundation models provide transferable pathology and CT representations that make such integration increasingly feasible \cite{uni,ctfm}. However, effective multimodal learning requires more than combining pretrained features. Global pooling can obscure tissue heterogeneity and regional anatomical information, while retaining thousands of image features complicates learning in small survival cohorts. Compact representations that preserve meaningful within-modality variation could provide a practical balance between expressivity and data efficiency. PANTHER addresses this challenge in histopathology by summarizing whole-slide images through morphological prototypes that encode tissue prevalence, appearance, and variability \cite{panther}. Multimodal prototyping extends this principle to histology and transcriptomics for survival prediction \cite{mmp}. Renal imaging studies have also explored probabilistic graph integration of radiology and pathology and feature fusion for recurrence prediction \cite{sparra,boeke}. These approaches motivate further investigation of how compact tissue representations can interact with spatially organized CT information.

We investigate bidirectional cross-attention between histology prototypes and kidney-centered CT tokens for disease-specific survival prediction as illustrated in Fig.\ref{fig:teaser}. The model retains prototype-specific tissue statistics and regional CT features, allowing each modality to condition the other before patient-level aggregation. We compare this formulation with unimodal, early-fusion, and late-fusion models on a shared TCGA-KIRC cohort and examine external histology-pathway transfer on CPTAC-CCRCC with CT withheld. Attribution analysis further connects predictions to modality contributions and image regions, supporting examination of complementary information and potential acquisition artifacts.



\section{Methods}

\subsection{Cohorts and endpoint}
The primary cohort included 210 TCGA-KIRC patients with histopathology images, CT scans, and disease-specific survival (DSS) labels from TCGA-CDR \cite{cdr}. The cohort included 46 DSS events with median follow-up of 3.7 years. Patients with missing or invalid outcomes or an empty mask were excluded, as were frozen sections. Imaging models used no clinical covariates. For external evaluation, we used PANTHER's CPTAC-CCRCC test list comprising 180 patients and 341 tumor slides, with 15 DSS events, 26 overall-survival (OS) events, and median follow-up of 2.5 years \cite{cptac}. Since only one DSS event occurred among the 52 patients with CT, external testing withheld CT. Adding 291 histology-only TCGA patients expanded training for this experiment to 501 patients with 109 DSS events. Slides were pooled per patient for external training and testing.

\subsection{Histology prototypes and anatomical CT tokens}
Frozen UNI2-h \cite{uni} encoded $256\times256$-pixel tissue patches at $20\times$ into 1,536-dimensional embeddings, yielding approximately 6,300 patches per slide. Following PANTHER \cite{panther}, we initialized 16 shared prototypes using $k$-means on 100,000 TCGA patches and fitted a patient-specific diagonal Gaussian mixture with one expectation--maximization iteration. Each prototype produced
\begin{equation}
\mathbf h_k=
[\log(\pi_k+10^{-4}),\boldsymbol\mu_k,\boldsymbol\sigma_k^2]
\in\mathbb R^{3073},
\end{equation}
encoding prevalence, mean appearance, and variability. Shared initialization established corresponding prototype slots across patients. The label-free dictionary was shared across TCGA folds and fixed for CPTAC.

For each patient, we selected one axial CT series from the earliest eligible study, prioritizing nephrographic or portal venous contrast phases. Eligible series contained $\geq20$ slices with spacing $\leq5$\,mm and coverage $\geq100$\,mm. Volumes were oriented superior--posterior--left, resampled to $3\times1\times1$\,mm, and scaled from $[-1024,2048]$ HU to $[0,1]$. BAMF masks defined a kidney-region box with a 30\,mm margin around retained components of $\geq10$\,ml within a 250\,mm superior--inferior span. Tumor labels were not supplied as predictive features. Frozen CT-FM \cite{ctfm} encoded the padded box in one pass. Adaptive pooling of its final 512-channel feature map onto a $2\times4\times4$ grid produced $M\leq32$ tokens $\mathbf c_j\in\mathbb R^{512}$, with centers expressed relative to the kidney-mask centroid in 100\,mm units. All feature standardization used training-fold statistics.

\begin{figure*}[h!]
    \centering    \includegraphics[width=\textwidth,height=0.25\textheight,keepaspectratio]{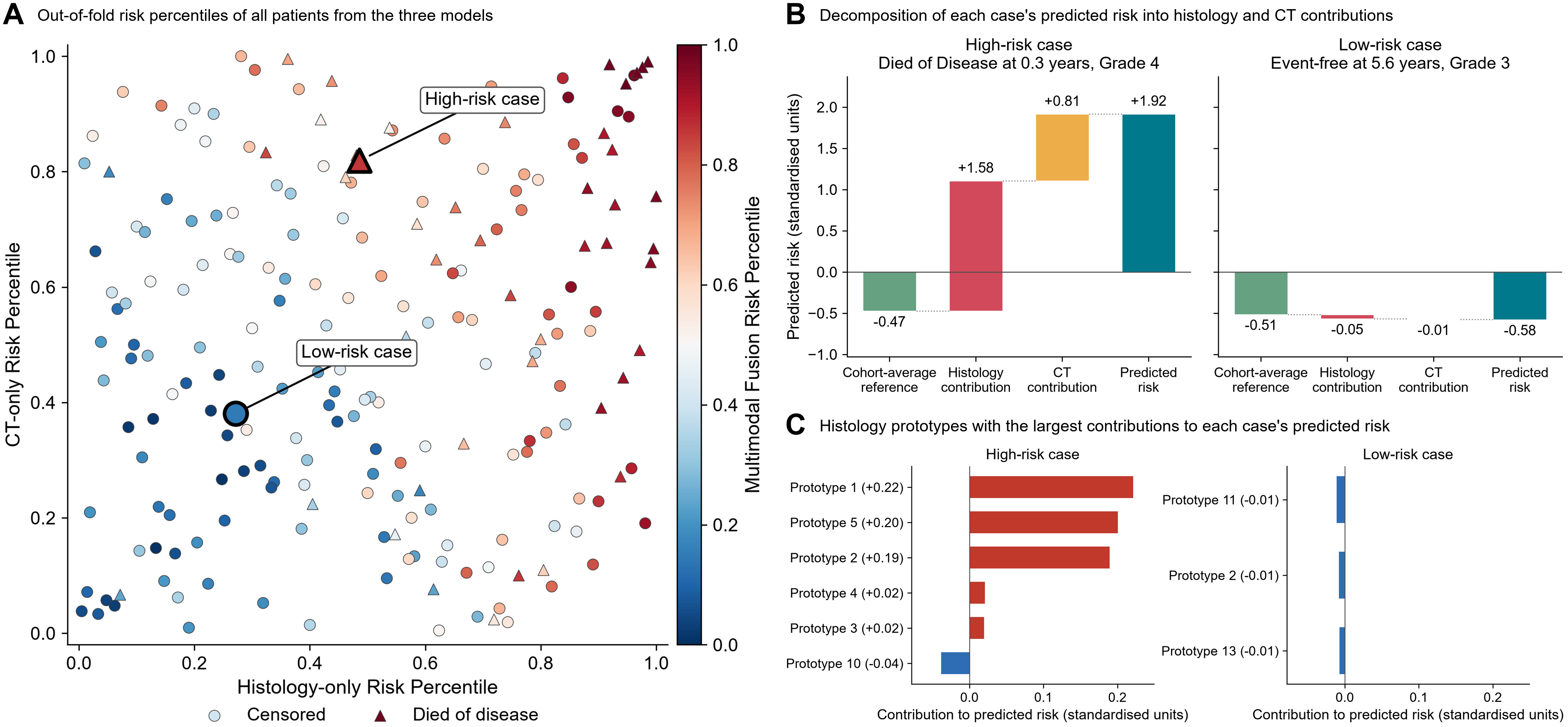}
    \caption{(A) Out-of-fold histology/CT risk percentiles, colored by fusion percentile; triangles denote DSS deaths. (B) In-sample IG decomposition into reference and modality contributions. (C) Prototype scores: red raises risk and blue lowers it.}
    \label{fig:ranking}
\end{figure*}

\subsection{Bidirectional fusion and survival learning}
We projected both token sets to $d=64$, adding learned prototype embeddings for histology and linearly embedded three-dimensional positions for CT. Each stream passed through one pre-normalized self-attention layer with four heads, feed-forward width 128, GELU activation, and dropout 0.25. Bidirectional cross-attention then updated the streams according to
\begin{equation}
X_m\leftarrow X_m+\operatorname{MHA}_{m\leftarrow n}
(\bar X_m,\bar X_n,\bar X_n),
\end{equation}
where $\bar X=\operatorname{LN}(X)$, LN denotes layer normalization, and MHA denotes multi-head attention. The pairs $(m,n)\in\{(h,c),(c,h)\}$ represent the two directions between histology ($h$) and CT ($c$), with separate directional weights and CT padding masks. Following residual feed-forward updates, mean-pooled histology and gated-attention-pooled CT \cite{abmil} were concatenated into a 128-dimensional representation and linearly mapped to risk $r_i$.

Training minimized the event-averaged negative Cox partial log-likelihood with Breslow handling of ties,
\begin{equation}
\mathcal L=-\frac{1}{N_e}\sum_{\ell}\left[
\sum_{i\in D_\ell}r_i-|D_\ell|\log\!
\sum_{j\,\mid\,T_j\geq t_\ell}e^{r_j}\right],
\end{equation}
where $D_\ell$ contains patients with events at distinct time $t_\ell$, $T_j$ is the observed follow-up time, and $N_e$ is the total event count. Training-only modality dropout withheld CT or histology with probability 0.15, disabling that modality's contribution to the other stream and zeroing its pooled representation. CT was withheld when unavailable, and both image encoders remained frozen.

\subsection{Comparators and evaluation}

We compared fusion against unimodal token models with linear heads and a histology MLP using log mixture weights and their weighted mean. Early fusion concatenated these histology features with pooled CT (2,064-D), followed by ridge Cox regression or a 32-unit MLP. Ridge penalties were tuned over $\{10^{-3},10^{-2},10^{-1},1,10\}$ using inner five-fold CV. Late fusion averaged standardized risks from independently trained token models. 

All models used identical event-stratified five-fold splits. Neural models trained for 40 full-batch epochs across three seeds with AdamW and gradient clipping at 1. Learning rates and weight decay were $(3\times10^{-4},10^{-2})$ for token models and $(10^{-4},0.1)$ for MLPs. Per-seed risks were standardized against training predictions before averaging. Design screening used a separate five-repeat, five-fold protocol.

We reported fold-wise Harrell's concordance (C-index) and three-year cumulative/dynamic AUROC and sensitivities at 80\% and 90\% specificity from pooled out-of-fold risks with inverse-probability-of-censoring weighting. Three-year events were cases, follow-up beyond three years defined controls, and earlier censorings were excluded (31 cases, 126 controls). Intervals used 1,000 patient-level bootstrap resamples. Thresholds derived from evaluation controls yielded descriptive ROC operating points.

\subsection{Attribution analysis}
Integrated gradients (IG) \cite{ig} explained the standardized three-seed risk from a model fitted to all 210 patients. The baseline replaced token content with cohort averages while retaining patient-specific CT positions and token count. A 64-step right-Riemann approximation yielded modality-, prototype-, and CT-token attributions with median absolute completeness error of 0.0026. To visualize histology contributions, prototype scores were distributed over patches using Gaussian-mixture posteriors and spatially aggregated. The resulting maps represent prototype projections rather than patch-level IG. Attributions were in-sample, whereas risk percentiles were out-of-fold.

\begin{table}[t]
\centering
\small
\setlength{\tabcolsep}{4pt}
\caption{DSS concordance comparison. Published histology-only
results are reproduced from PANTHER~\cite{panther}
(mean $\pm$ SD). Bold numbers indicate the highest reported
mean in each column. Published results use UNI and different
TCGA cohorts and splits, making comparisons contextual.}
\label{tab:panther_comparison}

\scalebox{0.80}{%
\begin{tabular}{@{}lcc@{}}
\toprule
Method & TCGA-KIRC & CPTAC-CCRCC \\
\midrule
ABMIL & $0.684\pm0.06$ & $0.613\pm0.06$ \\
TransMIL & $0.671\pm0.10$ & $0.639\pm0.04$ \\
DSMIL & $0.500\pm0.00$ & $0.500\pm0.00$ \\
AttnMISL & $0.649\pm0.09$ & $0.608\pm0.06$ \\
ILRA & $0.637\pm0.14$ & $0.611\pm0.03$ \\
DeepSets & $0.715\pm0.08$ & $0.634\pm0.01$ \\
ProtoCounts & $0.461\pm0.06$ & $0.503\pm0.09$ \\
H2T & $0.703\pm0.11$ & $0.631\pm0.04$ \\
OT & $0.695\pm0.09$ & $0.650\pm0.02$ \\
PANTHER Top + lin. & $0.741\pm0.11$ & $0.608\pm0.01$ \\
PANTHER Bot. + lin. & $0.524\pm0.14$ & $0.532\pm0.08$ \\
PANTHER WA + lin. & $0.730\pm0.07$ & $0.623\pm0.01$ \\
PANTHER All + lin. & $0.703\pm0.13$ & $0.649\pm0.04$ \\
PANTHER All + MLP & $0.716\pm0.10$ & $\mathbf{0.691\pm0.03}$ \\
\midrule
\textbf{Ours (fusion)}
& $\mathbf{0.760\pm0.081}$
& $0.629^{\dagger}$ \\
\bottomrule
\end{tabular}%
}

\vspace{3pt}
\begin{minipage}{\columnwidth}
\footnotesize
$^{\dagger}$CT withheld during external evaluation.
95\% bootstrap CI $[0.434,\,0.816]$.
\end{minipage}
\end{table}

\begin{table*}[t]
\centering\small
\caption{TCGA-KIRC DSS prediction ($n=210$). C-index is the 5-fold mean $\pm$ SD. 3-year metrics use censoring-adjusted, pooled out-of-fold predictions. Brackets indicate 95\% CIs and bold indicates highest means.}
\label{tab:performance}
\setlength{\tabcolsep}{7pt}
\scalebox{0.8}{%
\begin{tabular}{@{}lcccc@{}}
\toprule
Model & C-index & 3-year AUROC [95\% CI] & Sens., 80\% spec. & Sens., 90\% spec.\\
\midrule
CT tokens + positions & $0.671\pm0.130$ & 0.675 [0.564, 0.782] & 0.416 & 0.166\\
Histology tokens & $0.706\pm0.111$ & 0.723 [0.620, 0.825] & 0.570 & 0.443\\
Histology summary + MLP & $0.730\pm0.083$ & 0.718 [0.622, 0.814] & 0.448 & 0.418\\
Early fusion, ridge Cox & $0.688\pm0.115$ & 0.711 [0.610, 0.814] & 0.447 & 0.190\\
Early fusion, MLP & $0.744\pm0.110$ & 0.738 [0.639, 0.836] & 0.569 & \textbf{0.448}\\
Late fusion, equal risks & $0.736\pm0.116$ & 0.743 [0.640, 0.845] & 0.534 & 0.440\\
Cross-attention fusion (F2) & $\mathbf{0.760\pm0.081}$ & \textbf{0.773 [0.678, 0.860]} & \textbf{0.603} & 0.443\\
\bottomrule
\end{tabular}%
}
\end{table*}
\section{Results}
\begin{figure*}[t]
    \centering
    \includegraphics[width=\textwidth,height=0.3\textheight,keepaspectratio]{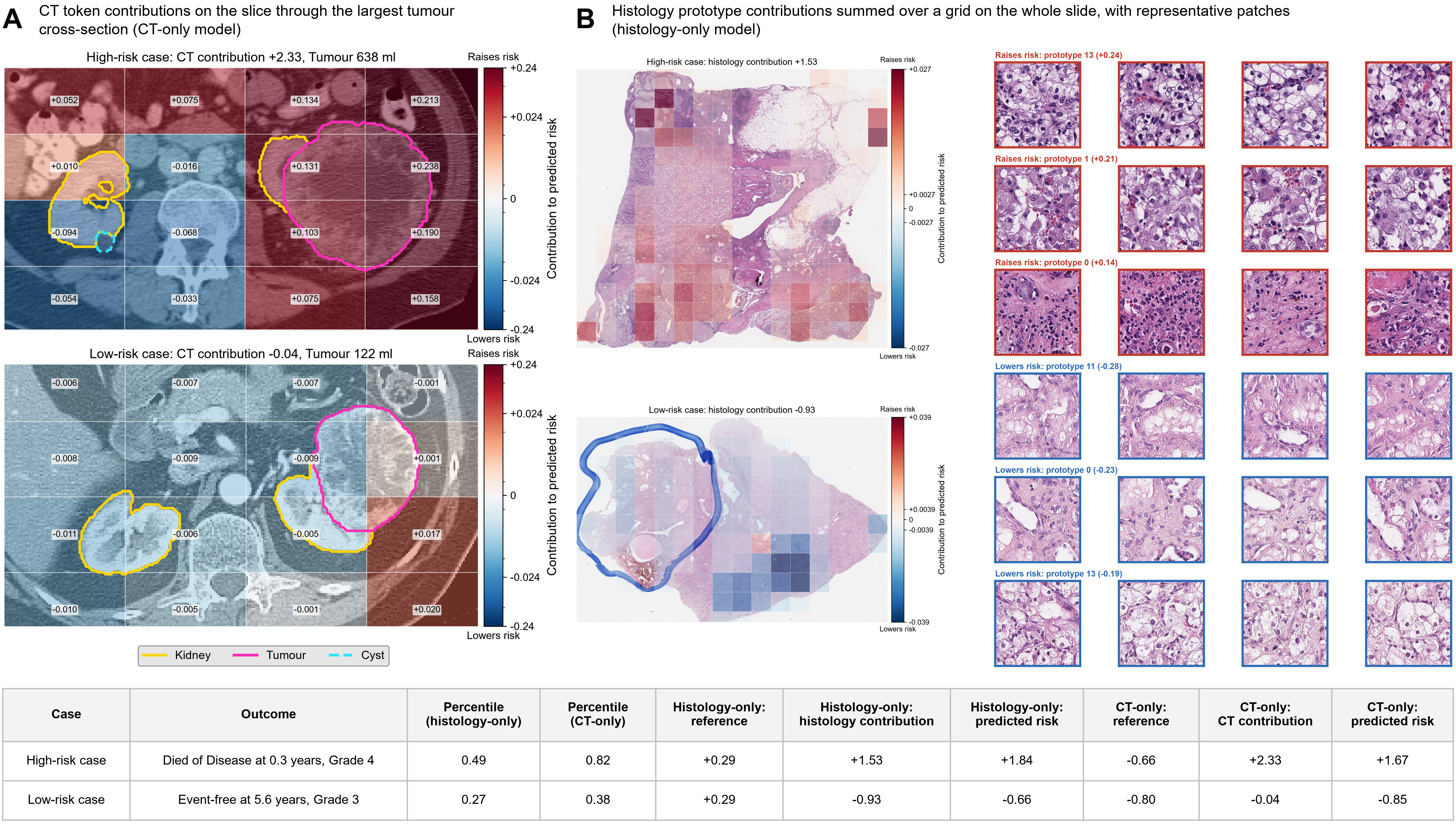}
    
    \caption{(A) CT token IG at the largest tumor cross-section, with shared scaling and kidney/tumor/cyst contours. (B) Prototype projections and patches with slide-specific scales. The blue ring is ink. Attributions are in-sample and percentiles out-of-fold. Small residuals reflect rounding and numerical integration.}
        \label{fig:unimodal}
\end{figure*}

\subsection{Fusion performance on a common cohort}
On the common cohort, cross-attention fusion achieved the highest mean C-index ($0.760\pm0.081$), three-year AUROC (0.773), and sensitivity at 80\% specificity (0.603), alongside the lowest C-index SD (Table~\ref{tab:performance}). At 90\% specificity, sensitivities were similar across nonlinear fusion designs (0.440--0.448). Although fusion increased mean concordance over histology MLP, histology tokens, and CT by 0.030, 0.054, and approximately 0.090, the corresponding paired intervals of $[-0.151,0.211]$, $[-0.104,0.212]$, and $[-0.086,0.265]$ included zero and did not establish superiority.

\subsection{External histology-pathway evaluation}
With CT withheld, the fusion model yielded external DSS C-index of 0.629 [0.434, 0.816], compared with 0.552 [0.339, 0.751] for histology tokens and 0.554 [0.403, 0.718] for histology MLP. This result followed expansion of TCGA training from 210 to 501 patients, which increased external DSS C-index from 0.541 to 0.629. Exploratory OS C-index was 0.674 [0.541, 0.789].

\subsection{From patient ranking to image evidence}
Fusion re-ranked two representative cases (Fig.~\ref{fig:ranking}A): a grade-4 patient who died at 0.3 years rose from the 49th (histology-only) and 82nd (CT-only) percentiles to the 86th, and a grade-3 patient event-free at 5.6 years fell from the 27th and 38th to the 15th. Fused risk rose from $-0.47$ to $+1.92$ through histology ($+1.58$) and CT ($+0.81$) in the high-risk case, while low-risk contributions were near zero ($-0.05$, $-0.01$; Fig.~\ref{fig:ranking}B). Unimodal models assigned larger CT ($+2.33$) and low-risk histology ($-0.93$) contributions (Fig.~\ref{fig:unimodal}), suggesting fusion redistributes evidence across modalities. Histology provided a median 77\% (IQR 62--87\%) of absolute cohort IG; CT contributions concentrated on the tumor side, and histology maps exposed potential pen/scanner shortcuts.

\section{Discussion}
Preserving within-modality structure while enabling cross-modal interaction yielded the highest mean DSS performance among evaluated designs. Although weaker alone, CT improved fusion performance beyond histology, highlighting its complementary value. Our approach integrates CT spatial information with histology prototype appearance and variability, extending cross-scale fusion \cite{panther,mmp,sparra,boeke}. PANTHER Top + linear achieved 0.741 on TCGA, while All + MLP achieved 0.691 on CPTAC, exceeding our CT-hidden external mean of 0.629 as seen from Table~\ref{tab:panther_comparison} . However, differences in cohorts, encoders, splits, and uncertainty estimates prevent direct superiority claims. Limited DSS events (46 internally and 15 externally) and wide intervals constrain inference. Differences in representations and training warrant matched cross-attention ablations. Screening-informed selection and a shared unlabeled dictionary may introduce optimism, while strong hospital signals in histology embeddings motivate site-held-out evaluation. Pen and scanner artifacts remained accessible to the model despite display filtering. Automatic masks and in-sample, baseline-dependent attributions further limit biological interpretation. External validation lacked CT, while calibration and incremental value beyond stage-inclusive clinical models remain untested. Future work should prioritize independent paired-modality validation, artifact removal, fold-specific prototypes, and matched ablations.


\section{Conclusion}
Cross-attention fusion achieved the highest average DSS performance among evaluated designs. Attribution revealed complementary image contributions and artifacts, motivating independent multimodal validation.

\section{Compliance with Ethical Standards}
This study used public, de-identified data, requiring no additional ethical approval.

\end{document}